\documentclass[12pt]{article}

\usepackage{graphicx,xcolor} 
\usepackage{bm}
\usepackage{array}
\usepackage{algorithm,setspace,booktabs,amsmath}
\usepackage{algpseudocode}
\usepackage[authoryear]{natbib}
\usepackage[margin=1in]{geometry}
\usepackage{authblk}

\title{Predictiveness Curve Assessment under Competing Risks for Risk Prediction Models}

\author[1]{Wei Tao}
\author[2]{Jing Ning}
\author[3]{Wen Li}
\author[1]{Wenyaw Chan}
\author[1]{Xi Luo}
\author[1]{Ruosha Li \thanks{Email: Ruosha.Li@uth.tmc.edu}}

\affil[1]{\small Department of Biostatistics and Data Science, The University of Texas Health Science Center at Houston, TX, USA.}
\affil[2]{\small Department of Biostatistics, The University of Texas MD Anderson Cancer Center, TX, USA.}
\affil[3]{\small Department of Internal Medicine, The University of Texas Health Science Center at Houston McGovern Medical School, TX, USA.}
\date{}

\begin{document}
\doublespacing

\newcommand{\BZ}{\boldsymbol{Z}}
\newcommand{\BB}{\boldsymbol{B}}
\newcommand{\Bbeta}{\bm{\beta}}
\newcommand{\Bhat}{\bm{\widehat\beta}}
\newcommand{\Btilde}{\widetilde\bm{\beta}}
\newcommand{\Btheta}{\bm{\theta}}

\maketitle

\abstract{The predictiveness curve is a valuable tool for predictive evaluation, risk stratification, and threshold selection in a target  population,  given a single biomarker or a prediction model. In the presence of competing risks, regression models are often used to generate predictive risk scores or probabilistic predictions targeting the cumulative incidence function—distinct from the  cumulative distribution function used in conventional predictiveness curve analyses . We propose estimation and inference procedures for the predictiveness curve with a competing risks regression model, to display the relationship between the cumulative incidence probability and the quantiles of model-based predictions. The estimation procedure combines cross-validation with a flexible regression model for $\tau-$year event risk given the model-based risk score, with corresponding inference procedures via perturbation resampling. The proposed methods perform satisfactorily in simulation studies and are implemented through an R package. We apply the proposed methods to a cirrhosis study to depict the predictiveness curve 
with model-based predictions for liver-related mortality.}\\

\noindent{\bf{Keywords:}} {Competing risks; Cumulative incidence; Fine and Gray model; Predictiveness curve; Liver mortality.}

\section{Introduction}\label{sec1}

Evaluating the predictive capacity of continuous markers or risk scores  is a crucial component in biomedical research \citep{moskowitz_quantifying_2004}. Using a continuous marker as an example, the predictiveness curve offers a comprehensive depiction of a marker's predictive capacity by plotting the risk level associated with each quantile of the marker's distribution \citep{huang_evaluating_2007}. It provides a direct illustration of the risk distribution in the population, demonstrates the effectiveness of the marker in risk stratification,
and aids in the selection of risk thresholds in medical decision making \citep{pepe_integrating_2007}. As such, the predictiveness curve provides a useful complement to other predictive metrics, such as the  receiver operating characteristic (ROC) curve.

Previous studies on the predictiveness curve mainly focused on binary outcomes. \cite{huang_evaluating_2007} introduced the predictiveness curve, with estimation and inference procedures under a flexible Box-Cox family.
The predictiveness curve was linked to several summary measures, such as the R-square statistic \cite[]{pepe_integrating_2007}, total gain \cite[]{bura2001binary}, and partial summary measures \cite[]{sachs_partial_2013}. A copula modeling procedure for curve estimation was proposed by \cite{escarela_copula_2020} through constructing the joint density of the marker and the outcome. For survival outcomes, \cite{viallon2011discrimination} studied the relationship between the predictiveness curve and the area under the ROC curve, and \cite{escarela2023copula} considered  curve estimation based on parametric and semi-parametric copula modeling.

Competing risks occur when an individual is at risk of multiple failure events, and the occurrence of one event prevents the occurrence of other events. This scenario is commonly encountered in medical research. For instance, in a study involving adults with cirrhosis undergoing transjugular intrahepatic portosystemic shunt (TIPS), patients may experience liver-related mortality, mortality from other causes, or liver transplantation \cite[]{vizzutti_mortality_2023}, where these events form a competing risks structure. Thus, analysis and prediction for the main event of interest, liver-related mortality, needs to account for the presence of the other events.

 Several regression methods for competing risks have been developed to facilitate the estimation and prediction of the cumulative incidence function (CIF), which quantifies the probability of a specific event occurring in the presence of other competing events. The Fine and Gray model is  one of the most  widely used approach. It extends the Cox proportional hazards model to model the subdistribution hazard function, thereby directly relating the CIF to covariates \cite{fine_proportional_1999}.  \cite{klein_regression_2005} introduced the pseudo-observation approach for cumulative incidence regression. Additionally, \cite{jeong2006direct} studied parametric regression of the cumulative incidence through a Gompertz distribution,
\cite{scheike2008predicting} proposed a direct binomial regression model to capture time-varying effects of covariates on the CIF, while \cite{bellach_weighted_2019} employed nonparametric maximum likelihood estimation techniques. These models enable predictive risk scores as combinations of multiple risk factors, as well as probabilistic predictions for the CIF.

When a competing risks regression model is applied for risk prediction and stratification, it is important to assess its predictive capacity in the population and understand how well it stratifies risk.
Our study aims to estimate the predictiveness curve given a competing risks regression model, to explicitly display how the risk level of the event of interest
evolve with the distribution of the predicted value. In doing this, we allow the prediction model to be subject to model misspecification, by treating it as a working model instead of a true model. Moreover, our inference procedures account for the additional variability due to the estimated model parameters. The proposed method is implemented in an R package \textit{cmpCurve} ({https://github.com/rli1010/cmpCurve}) to support practical applications.

The rest of this paper is organized as follows. Section 2 introduces definition and estimation of the predictiveness curve for a competing risks regression model, coupled with the variance estimation procedure. Section 3 demonstrates the performance of the proposed methods in simulation studies under different settings and sample sizes. In Section 4, we apply the proposed methods to a real-world dataset on cirrhosis to evaluate the predictiveness curve with a predictive model for
liver-related mortality. We conclude with some discussions in Section 5.

\section{Method}\label{sec2}

\subsection{Predictiveness Curve
for a Working Competing Risks Prediction Model}

Consider a scenario where each individual is subject to failure from one of 
$K$ competing risks.  Let $T$ denote the failure time of the earliest event, and  $\epsilon\ \in \{1 , ..., K\}$ denote the cause of failure. The failure time $T$ might be censored by an independent random variable $C$.  Define the observed time as $Y = \min(T,C)$ and the censoring indicator as $\Delta = I(T \leq C)$. Additionally, we have a $d$-dimensional vector of baseline covariates represented by $\BZ = (Z_1, Z_2, ..., Z_d) ^\top$.  We observe $n$ independently and identically distributed replicates of the data $\{Y, \Delta \epsilon, \BZ\}$, where $\Delta \epsilon\in\{0,1,2,...,K\}$.
Without loss of generality, we focus on the failure associated with Cause 1. Let $\tau$ denote a particular time point of interest, which can be determined based on clinical relevance. An individual's status for the cause-1 event of interest can be expressed as $I(T_i \leq \tau, \epsilon_i\ = 1)$, whose probability is given by the cumulative incidence function (CIF), $F_1(\tau; \BZ_i) = P(T_i \leq \tau, \epsilon_i\ = 1 \mid \BZ_i)$. 

Competing risks regression models can be used generate predictive risk scores or predicted cumulative incidence probabilities. For example, the Fine and Gray model \citep{fine_proportional_1999} assumes that
\begin{equation}\label{eqn:FGmodel}
\lambda_1(t; \BZ) = \lambda_{10}(t) \exp(\bm{\beta}^\top \BZ ),
\end{equation}
where $\lambda_1(t; \BZ)$ denotes the conditional sub-distribution hazard function, $\lambda_{10}(t)$ denotes an unspecified baseline sub-distribution hazard function, and $\bm{\beta}$ represents the vector of unknown regression coefficients. Fitting the model on observed data provides an estimated coefficient $\widehat{\bm{\beta}}$.
While we focus on  $\widehat{\bm{\beta}}$ from the Fine and Gray model for illustrative purposes in the work, our methods below can also accommodate predictors from other competing risks regression models, such as \cite{jeong2006direct} and \cite{scheike2008predicting}.

Based on a fitted regression model, it is natural to stratify patients according to their risks of the cause-1 event using 
$\xi(\BZ_i, \widehat{\bm{\beta}})$, where 
$\xi(\BZ, {\bm{\beta}})=\BZ^\top \bm{\Bbeta}$. We allow the regression model in \eqref{eqn:FGmodel} to be subject to potential model misspecification by treating  it as a working model. Despite the potential model misspecification, it has been shown  that $\widehat{\bm{\beta}}$ converges in probability to deterministic value, denoted as $\widetilde{\bm{\beta}}$, as the sample size increase \citep{ding_evaluation_2021}. Thus, when $\max_i ||Z_i||$ is bounded, we anticipate that $\xi(\BZ, \widehat{\bm{\beta}})=\BZ^\top\Bhat$ converges uniformly in $\BZ$ to a deterministic function, denoted as 
$\xi(\BZ, \widetilde{\Bbeta})=\BZ^\top \widetilde{\Bbeta}$.  
We aim to evaluate the performance of the model-based risk score $\xi(\BZ, \widetilde{\bm{\beta}})$ in predicting the $\tau$-year outcome. While we use the predictive risk score here, the predictiveness curve for model-based CIF  predictions would be equivalent, as the predicted CIFs have a monotone relationship with the predictive risk scores, and the predictiveness curve is defined  based on the quantiles of the predicted value.

Denote $\xi_i=\BZ_i^\top \widetilde{\Bbeta}$ for notational simplicity below, and let $Q(v)$ denote the $v_{th}$ quantile of this model-based risk score. We extend the predictiveness curve definition for a binary outcome \citep{huang_evaluating_2007} to the binary indicator of $I(T_i\leq \tau,\varepsilon_i=1)$ under competing risks data for a model-based risk score. Specifically, let $R(v)$ represent the $\tau-$year CIF associated with the $v$th quantile of the risk score, expressed as
\begin{equation}\label{eqn:rvdef}
R(v) = P\{T\leq \tau, \epsilon\ = 1 \mid \xi = Q(v)\},\quad v\in (0,1).
\end{equation}
We aim to estimate the competing risks predictiveness curve, which plots $R(v)$ against $v$. As a special case, we can also accommodate the situation of a single biomarker by letting $\xi_i=Z_{1i}$, assuming without loss of generality that higher values of this biomarker correspond to worse outcomes.

The predictiveness curve $R(v)$ offers a comprehensive visualization based on the CIF, providing direct insights into  how the risk level evolves with the quantile of a prognostic biomarker or model-based risk score in the population.
It allows for evaluating the effectiveness of different markers and risk scores in risk stratification. A steeper curve indicates greater variation in predicted risk across the distribution of the score, reflecting better  discrimination between high- and low-risk individuals by the score. Additionally,  scores can be evaluated based on the proportions of the population classified into clinically relevant risk categories (e.g., low risk, high risk, intermediate risk) using predefined thresholds. 

\begin{figure*}[ht]
\centerline{\includegraphics[width=0.8\textwidth]{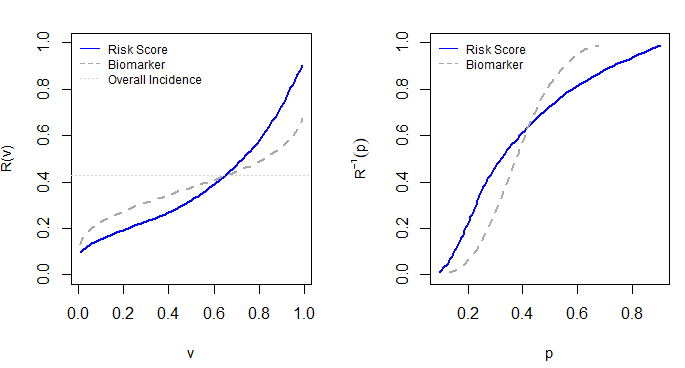}}
\caption{Illustrative Example of the Predictiveness Curve $R(v)$ (left panel) and its inverse function $R^{-1}(p)$ (right panel)
    for a Single Biomarker or a Risk Score in the Competing Risks Setting.\label{fig1}}
\end{figure*}

Figure 1 provides an illustrative example of the predictiveness curves for a single biomarker and a predictive risk score, respectively. By utilizing the quantile scale on the x-axis, the risk score and  biomarker can be mapped to a common scale.  From the left panel, we observe that the cumulative incidence risk at the 90th percentile (i.e., R(0.9)) is 0.73 for the risk score, compared to 0.56 for the single biomarker. Thus, patient in the top 10\% of the risk score distribution are at greater risk compared to those in the top 10\% by the single biomarker. In our example, the blue curve (risk score) has a steeper shape than the gray curve (single biomarker), indicating superior risk stratification performance.

An alternative perspective is offered by examining its inverse function. For a specific risk level $p$, $R^{-1}(p)$ represents the proportion of the population with cumulative incidence probability less than $p$ according to the risk score. 
Suppose that $p_l$ and $p_h$ define the thresholds for  ``low risk" and  ``high risk".  Then the portions of the population that fall into low, high, and intermediate risk categories, are represented by $R^{-1}(p_l)$, $1-R^{-1}(p_h)$, and $R^{-1}(p_h)-R^{-1}(p_l)$ respectively. Taking the  risk score in Figure 1 as an example, suppose that $p_h=0.25$ and $p_l=0.5$ for this particular disease, then it can be indicative of low risk for 36\% of the population (as calculated by $R^{-1}(0.25)$), of high risk for 27\% of the population (determined by $1-R^{-1}(0.5)$), leaving 37\% of patients within the intermediate risk range.

\subsection{Estimation Procedures}

As $\widetilde{\Bbeta}$ is unknown, we began by estimating it under a working Fine and Gray model. We employed a cross-validation (CV) scheme to alleviate the potential of optimism. Specifically, we partitioned the observed data into two subsets, denoted as  $\mathcal{D}_A$ and $\mathcal{D}_B$, which were then utilized for estimating the risk scores and the predictiveness curve, respectively. 
Without loss of generality, we opted for a two-fold repeated cross-validation strategy. The estimated predictiveness curve derived from the initial  split is represented as $\hat{R}(v)^{(1)}$. We then interchanged the roles of  $\mathcal{D}_A$ and $\mathcal{D}_B$, subsequently conducting another round of estimation to yield $\hat{R}(v)^{(2)}$. The cross-validation-type estimator is then defined as  the average of the two:
\[
\hat{R}(v)^{(CV)} = \left\{ \hat{R}(v)^{(1)} + \hat{R}(v)^{(2)} \right\} / 2.
\]
To improve finite-sample stability, this cross-validation was repeatedly performed multiple times, and  the final estimate was obtained by calculating the average across all repetitions. 

We next outline the detailed estimation procedures for $\hat{R}(v)$ using a specific split
$\{\mathcal{D}_A,\mathcal{D}_B\}$, with sample size $(n_A, n_B)$ respectively, where $n=n_A+n_B$. After fitting a Fine and Gray model on \( \mathcal{D}_{A} \), we obtain the corresponding coefficient $\Bhat$.
Consequently, for an individual in the test set $\mathcal{D}_B$, the model-based risk score ${\xi}_i
 = \widetilde{\bm{\beta}}^\top \mathbf{Z}_i$
can be estimated by $\widehat{\xi}_i
 = \Bhat^\top \mathbf{Z}_i$. 
 
  Instead of directly adopting the model-based CIF according to the Fine and Gray model in \eqref{eqn:FGmodel}, which considers a constant coefficient over time, we consider a more flexible binomial model for the CIF locally at time $\tau$ given $\xi_i$ as
\begin{equation}\label{eqn:CIFmodel}
P(T_i \leq \tau, \epsilon_i = 1 \mid \xi_i) =  g\{\theta_0+{\Btheta_B^\top} \widetilde{\bm{B}}(\xi_i)\}. \end{equation}
Here $g(\cdot)$ denotes an increasing link function, such as the inverse-logit link or inverse-probit link. We use the inverse-logit link below. 
$\widetilde{\bm{B}}(\xi)=\{q_1(\xi), \dots, q_m(\xi)\}^\top$ include prespecified basis functions, where $m$ is a small integer.  In this work, we adopt the restricted cubic spline (RCS) basis functions with a prespecified number of knots. This allows for non-linear relationships between the model-based risk score $\xi_i$ and the
CIF. In practice, $3-5$ knots are typically sufficient for representing the underlying relationships \citep{stone1986generalized,harrell2001regression}. When $Q$ knots are specified, $\widetilde{\bm{B}}(\cdot)$ includes 1 linear term plus $Q-2$ additional terms, such that $m=Q-1$.
Next, $\bm\theta_B$ is the corresponding coefficient with the same dimension as $\widetilde{\bm{B}}(\xi)$. We denote $\bm\theta=(\theta_0,\bm\theta_B^\top)^\top$ and $\bm{B}(\xi)=\{1,\widetilde{\bm{B}}(\xi)^\top\}^\top$ below.

In the presence of right censoring,  we propose the following weighted objective function for estimating $\Btheta$ based on the $n_B$ observations in $\mathcal{D}_B$:
\begin{equation}
l(\bm\theta) = \sum_{i=1}^{n_B} \frac{I(T_i \land \tau \leq C_i)}{\hat{G}(Y_i \land \tau\mid\BZ_i)}
\log\bigg[\bigg\{\frac{e^{\{\bm{\theta}^\top \BB(\hat\xi_i)\}}}{1 + e^{\{\bm{\theta}^\top \BB(\hat\xi_i)\}}}\bigg\}
^{\delta_i} \bigg\{\frac{1}{1 + e^{\{\bm{\theta}^\top \BB(\hat\xi_i)\}}}\bigg\}^{1-\delta_i}\bigg].
\end{equation}
Here, $ \delta_i = I(Y_i \leq \tau, \Delta_i\epsilon_i = 1) $  is an indicator variable of observed cause-1 event by time $\tau$, and $I(T_i \land \tau \leq C_i)= I(Y_i \leq \tau) \Delta_i  + I(Y_i \geq \tau)$ equals 1 if the value of $I(T_i\leq\tau,\varepsilon_i=1)$ can be determined from the observed data and 0 otherwise. The $\hat{G}(Y_i \land \tau\mid\BZ_i)$ is the corresponding IPCW weight, where $\hat{G}(t\mid\BZ)$ is an estimator of $P(C\geq t|\BZ)$ \citep{li_estimating_2011}. In situations where $C$ is random and independent of both outcome and covariates, we can estimate $G(\cdot)$ using the Kaplan-Meier estimator. Otherwise, we can adopt a parametric or semi-parametric regression model for $C$ given $\BZ$ under the conditional independent censoring. To maximize the objective function, we can adapt the existing function for logistic regression with RCS splines to incorporate the first term as weights.
This can be efficiently achieved by employing the \verb|lrm()| function from the R package \verb|rms| \citep{rmsmanual}. We denote the corresponding maximizer as $\widehat{\Btheta}$.

We next calculate the $v$th quantile of the risk score $\widehat{Q}(v)$ by taking the empirical quantile of $\hat\xi_i$, $i=1,2,...,n_B$, for the observations in $\mathcal{D}_B$. These procedures are then combined under model \eqref{eqn:CIFmodel} to obtain the predictiveness curve estimate as 
\[\hat{R}(v)^{(1)} = g\big[\bm{\widehat\theta}^\top \bm{B}\{\widehat{Q}(v)\}\big], \quad v\in [\nu_0,\nu_1].\]
 Here,
$\nu_0$ and $\nu_1$ are constants close to $0$ and $1$,
respectively, but do not contain the boundary regions.  This  avoids the inherent  instability with both $\hat{Q}(v)$ and the RCS estimation at the tails. In practice, we can estimate the curve over a finely spaced grid with a small grid size such as $0.01$. We then repeat the procedure to obtain $\hat{R}(v)^{(2)}$ and the final CV-type estimator $\hat{R}(v)^{(CV)}$. The estimation can be implemented as a special case of Algorithm \ref{eqn:alg} below by ignoring the perturbation weights.
The corresponding R package \textit{cmpCurve} is available at {https://github.com/rli1010/cmpCurve}.

\subsection{Variance Estimation}

The estimation of variance for the proposed estimator is complicated by the added variability arising from the estimated $\Bhat$ and the implementation of the repeated cross-validation scheme. To approximate the distribution of the proposed estimator, we developed a perturbed-resampling procedures to account for all sources of variance comprehensively (\cite{jin2001simple,ding_evaluation_2021}). With a large integer $E$, such as $E$ = 400, let $\boldsymbol{\omega}^{(e)} = (\omega_1^{(e)}, \omega_2^{(e)}, \ldots, \omega_n^{(e)})$, where $e = 1, \ldots, E$ and $n$ is the total sample size.
This represents a matrix of dimensions $n \times E$, where each element is an independent copy of a random variable drawn from a unit exponential distribution. First, we obtain the perturbed estimate of $\bm\beta$ by fitting the Fine \& Gray model incorporating the perturbation weights as sampling weights in the training set, and then we have the perturbed risk score $\hat{\xi}_i^{(e)} = \widehat{\bm\beta}^{(e)\top} \BZ_{i}$. To account for the variability of $\widehat{\bm\theta}$, we adopt the perturbed objective function as
\begin{equation}
l^{(e)}(\bm\theta) = \sum_{i=1}^{n_B} \frac{\omega_{i}^{(e)}I(Y_i \land \tau \leq C_i)}{\hat{G}^{(e)}(Y_i \land \tau\mid\BZ_i)}
\log\bigg[\bigg\{\frac{e^{\{\bm{\theta}^\top \BB(\hat\xi_i^{(e)})\}}}{1 + e^{\{\bm{\theta}^\top \BB(\hat\xi_i^{(e)})\}}}\bigg\}
^{\delta_i} \bigg\{\frac{1}{1 + e^{\{\bm{\theta}^\top \BB(\hat\xi_i^{(e)})\}}}\bigg\}^{1-\delta_i}\bigg].
\end{equation}
where $\hat{G}^{(e)}(Y_i \land \tau)$ is the perturbed Kaplan-Meier estimator for ${G}(Y_i \land \tau)$. We also obtain a perturbed $\hat{Q}(v)^{(e)}$ by adopting the weighted quantile function in R, which is efficiently computed using the \verb|wtd.quantil()| function from the R package  \verb|Hmisc|. Finally, we derive the perturbed risk probability:
\begin{equation}
\hat{R}(v)^{(e,1)} = g[\hat{\bm{\theta}}^{{(e)}\top} \bm{B}\{\hat{Q}^{(e)}(v)\}]
\end{equation}
The proposed procedure for estimating variance is outlined in Algorithm 1. 

\begin{algorithm}
\caption{Variance Estimation Procedure}
\label{eqn:alg}
\begin{enumerate}
    \item Generate the perturbation weight $\boldsymbol{\omega}^{(e)} = (\omega_1^{(e)}, \omega_2^{(e)}, \ldots, \omega_n^{(e)}), e = 1, \ldots, E$, by sampling from the unit exponential distribution.
\item For {$e = 1$ \textbf{to} $E$:}
\begin{enumerate}
    \item  Fit the Fine and Gray model to the training set $\mathcal{D}_A$, incorporating the weights, to obtain $\Bhat^{(e)}$.
     \item For each subject in the test set $\mathcal{D}_B$, compute  $\hat{\xi}_i^{(e)} = \hat{\bm\beta}^{(e)\top} \BZ_{i}$.
     \item Use $\hat{\xi_i}^{(e)}$ along with perturbation weight to calculate the empirical quantile, denoted as ${\hat{Q}(v)}^{(e)}$. 
     \item Use the perturbed $l^{(e)}(\theta)$ to estimate the perturbed $\hat{\bm{\theta}}^{(e)}$.
     \item Plug $\hat{\bm{\theta}}^{(e)}$ and ${\hat{Q}(v)}^{(e)}$ into
$\hat{R}(v)^{(e,1)} = g[\hat{\bm{\theta}}^{{(e)}\top} \bm{B}\{\hat{Q}^{(e)}(v)\}]$.

     \item Switch $\mathcal{D}_A$ and $\mathcal{D}_B$, repeat steps (a)-(e) to obtain $\hat{R}(v)^{(e,2)}$.
     \item Calculate $\hat{R}(v)^{(e,CV)} = \{ \hat{R}(v)^{(e,1)} + \hat{R}(v)^{(e,2)} \} / 2$.
     \item Perform the two-fold cross-validation procedure repeatedly and calculate the average of the results, denoted by $\hat{R}(v)^{(e)}$.
\end{enumerate}

\item Estimate the variance of $\hat{R}(v)$ using the sample variance of $\hat{R}(v)^{(e)}$, $e = 1, 2, \ldots, E$.
\end{enumerate}
\end{algorithm}

\section{Simulation Study}
\subsection{Data simulation}
Our simulation Setting 1 mimics that in \cite{fine_proportional_1999}. We started by generating covariates $\BZ_i = (Z_{i1}, Z_{i2})$, with each element $Z_{ij}$ drawn from a standard normal distribution for $j = 1, 2$. The CIF for Cause 1 failure follows
\begin{equation*}
P(T_i\leq t, \epsilon_i= 1 \mid Z_i) = 1 - [1 - \gamma \{1 - \exp(-t/3) \}]^{\exp(Z_{i1}\beta_{11} + Z_{i2}\beta_{12})}.
\end{equation*}
The true parameter values for $(\beta_{11}, \beta_{12}, \beta_{21}, \beta_{22})$ were set to $(0.5, 0.5, -0.5, 0.5)$, and $\gamma$ was set at 0.48. We generated censoring time $C \sim 4.2 \times \text{Beta}(5, 1)$, such that the independent censoring rate was 30\%. 

For Setting 2, we considered two covariates, $Z_1\sim Bernoulli(0.5)$ and $Z_2\sim Normal(0,1)$. Given $\BZ_i$, the cause indicator $\varepsilon_i$ was set to 1 with probability
$0.75\exp(Z_{1i}+Z_{2i})/\{1+\exp(Z_{1i}+Z_{2i})\}$, and 2 otherwise. We then generated $T\mid \varepsilon_i=1$
from $c_1\cdot \text{Weibull}\{shape=2,scale=\exp(-0.5 Z_{1i}-0.75 Z_{2i})\}$
and $T\mid \varepsilon_i=2$ from a Uniform(0,5.6) distribution.
We generated the independent censoring time $C \sim 4.3 \times \text{Beta}(5, 1)$ and adjusted the constant $c_1$ to obtain an independent censoring rate of $30\%$. The Fine and Gray model is correctly specified in Setting 1 but mis-specified in Setting 2, where it serves as a working prediction model. The observed proportion of cause-1 failure was approximately 36\% under Setting 1 and 35\% under Setting 2.
We set $\tau = 4$ under both settings.
It is worth noting that our flexible binomial model \eqref{eqn:CIFmodel} for estimating the predictiveness curve is also subject to mis-specificaton under both settings, but the spline parameterization is anticipated to provide adequate approximation.

\begin{figure}[h]
\centerline{\includegraphics[width=0.7\textwidth]{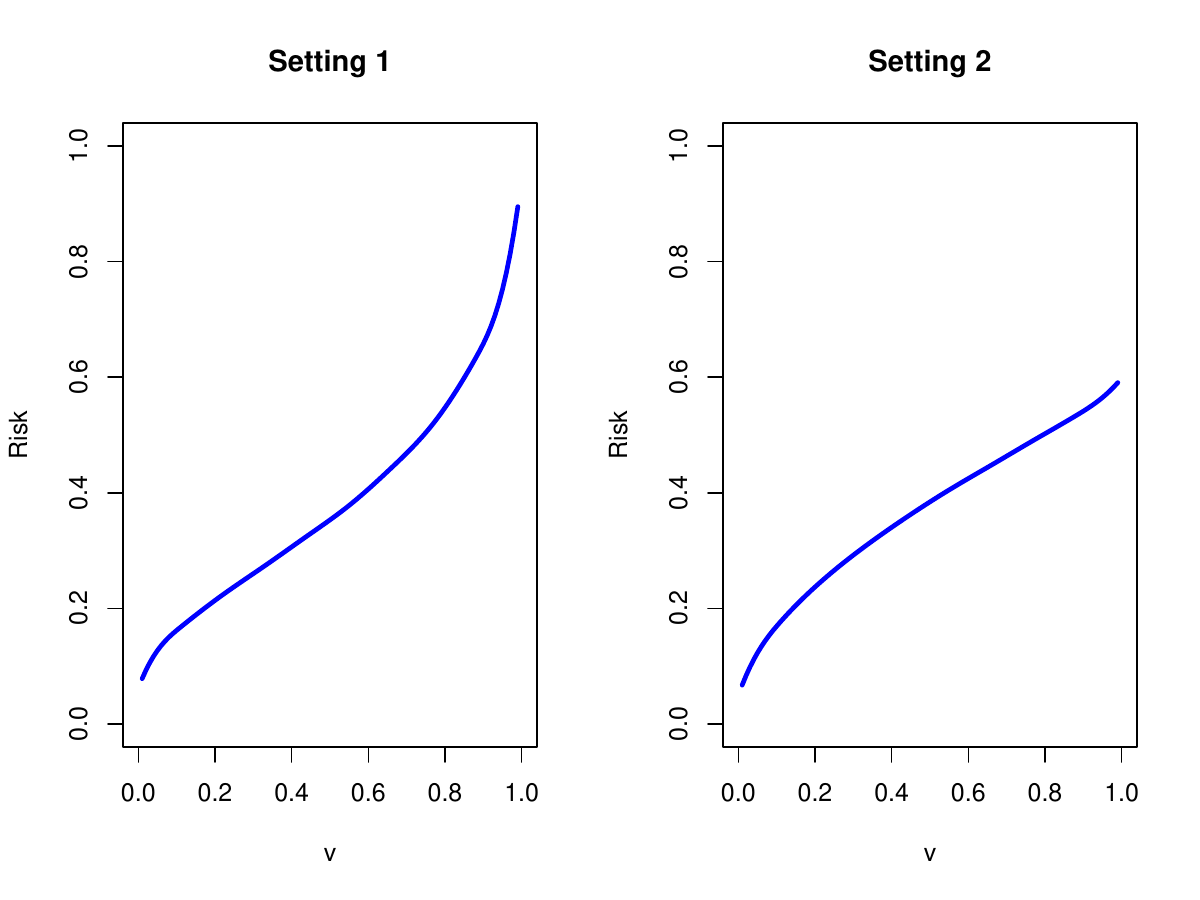}}
\caption{True Predictiveness Curves under Simulation Settings 1 \& 2.\label{fig2}}
\end{figure}

Figure 2 presents the corresponding true predictiveness curves. The true curve has an explicit form for Setting 1. For Setting 2, we first estimated the limiting $\widetilde{\Bbeta}$
with a huge dataset. We then approximated the true curve using another huge uncensored dataset, repeating the process many times to compute an average.
The shapes of the curves are different under the two settings. Under Setting 2, the maximum of predictiveness curve stays further away from $1$ when compared to that of Setting 1.

\subsection{Simulation result}
We implemented the proposed methods with sample sizes of $n=200$, $400$, $800$ for each setting on $1,000$ simulated datasets.  
For each simulated dataset, the cross-validation process was repeated five times. Standard error calculations were derived using perturbation with $E = 400$. Wald-type confidence intervals were constructed based on logit-transformed $\hat{R}(v)$ and its standard error. For the purpose of comparison, we also implemented a simpler version of the proposed method, by only including the linear term of $\xi_i$ without additional spline terms in \eqref{eqn:CIFmodel}. We refer to the two parameterization as RCS (with splines) and GLM (with only linear term) below.

Table 1 displays the results for $\hat{R}(v)$ using the two different parameterizations. The bias in estimates across all sample sizes ($n = 200, 400, 800$) was minimal. The empirical standard error (ESE) and the average standard error (ASE) were consistent across various sample sizes, both showing a decreasing trend as sample sizes increased. The coverage probability (CP) of 95\% confidence intervals for $R(v)$ was close to the nominal level in most scenarios, particularly for larger sample sizes, suggesting adequate performance of the inference procedure.  The RCS estimator tended to have slightly a larger standard error than the basic parameterization but showed some advantage in terms of bias and  CP, especially for Setting 2 at larger sample sizes. Thus, a more flexible spline parameterization is preferred for larger sample sizes, while a basic parameterization may be adequate for small sample sizes.

Table 2 provides simulation results for the corresponding $\hat{R}^{-1}(p)$, which reflects the estimated proportion of patients with risk less than $p$ according to the model-based risk score. In general, the proposed method based on the RCS parameterization showed small bias, with the bias shrinking towards 0 as sample size increased. The bias was slightly larger when $n=200$
for Setting 2 with $p=0.5$, which was close to the maximum of $R(v)$ in Figure 2. The empirical coverage rates were also close to the nominal level. By comparison, the basic parameterization  led to larger bias that did not fully diminish with increasing sample size.

\begin{center}
\begin{table*}[h]%
\caption{Simulation results for $\hat{R}(v)$.\label{tab1}}
 \begin{tabular}{lcccccccccc}
    \hline\hline
    \\
            \multicolumn{2}{l}{Setting 1}& \multicolumn{4}{c}{GLM}  &     & \multicolumn{4}{c}{RCS} \\
      \multicolumn{2}{c}{} & $v=0.1$ & $v=0.3$ & $v=0.5$ & $v=0.7$ &  & $v=0.1$ & $v=0.3$ & $v=0.5$ & $v=0.7$ \\
      \hline
      \\
      \multicolumn{2}{l}{True $R(v)$} & 0.162 & 0.260 & 0.353 & 0.469 & & 0.162 &0.260 & 0.353 & 0.469 \\\addlinespace
      {Bias}& $n=200$ & 0.000 & 0.007 & 0.018 & 0.023 & & 0.015 & -0.004 & 0.001 & 0.020 \\
            & $n=400$ & -0.009 & 0.002 & 0.014 & 0.019 & & 0.006 & -0.006 & -0.004 & 0.011 \\
            & $n=800$ & -0.010 & 0.002 & 0.013 & 0.016 & & 0.004 & -0.005 & -0.004 & 0.006 \\
            \addlinespace
      {ESE}& $n=200$ & 0.053 & 0.049 & 0.047 & 0.059 & & 0.061 & 0.052 & 0.062 & 0.066 \\
            & $n=400$ & 0.035 & 0.034 & 0.033 & 0.042 & & 0.041 &  0.035 & 0.043 & 0.046 \\
            & $n=800$ & 0.025 & 0.024 & 0.023 & 0.030 &  &0.029 & 0.025 & 0.031 & 0.032 \\\addlinespace
      {ASE}& $n=200$ & 0.052 & 0.047 & 0.046 & 0.056 & & 0.058 & 0.050 & 0.058 & 0.063 \\
            & $n=400$ & 0.035 & 0.033 & 0.032 & 0.040 & & 0.040 & 0.034 & 0.042 & 0.045 \\
            & $n=800$ & 0.025 & 0.024 & 0.023 & 0.029 & & 0.028 & 0.024 & 0.030 & 0.032 \\\addlinespace
      {CP}& $n=200$ & 0.945 & 0.942 & 0.930 & 0.937 & & 0.928 & 0.946 & 0.938 & 0.941 \\
            & $n=400$ & 0.941 & 0.950 & 0.923 & 0.926 & & 0.943 & 0.947 & 0.942 & 0.946 \\
            & $n=800$ & 0.931 & 0.946 & 0.908 & 0.915 & & 0.937 & 0.944 & 0.946 & 0.946 \\
  & & & & & & & & &\\
            \multicolumn{2}{l}{Setting 2}& \multicolumn{4}{c}{GLM}   &    & \multicolumn{4}{c}{RCS} \\
      \multicolumn{2}{l}{} & $v=0.1$ & $v=0.3$ & $v=0.5$ & $v=0.7$ & & $v=0.1$ & $v=0.3$ & $v=0.5$ & $v=0.7$ \\
      \hline
      \\
      \multicolumn{2}{l}{True $R(v)$} & 0.169 & 0.292 & 0.384 & 0.463 & & 0.169 & 0.292 & 0.384 & 0.463 \\\addlinespace
      {Bias}& $n=200$ & 0.023 & -0.013 & -0.023 & -0.010 & & -0.001 & -0.010 & 0.003 & 0.012 \\
            & $n=400$ & 0.030 & -0.008 & -0.024 & -0.019 & & 0.004 & -0.002 & 0.003 & 0.002 \\
            & $n=800$ & 0.034 & -0.006 & -0.026 & -0.026 & & 0.006 & 0.002 & 0.003 & -0.004 \\\addlinespace
      {ESE}& $n=200$ & 0.051 & 0.046 & 0.043 & 0.050 & & 0.057 & 0.052 & 0.058 & 0.054 \\
            & $n=400$ & 0.037 & 0.032 & 0.030 & 0.035 & & 0.042 & 0.036 & 0.042 & 0.039 \\
            & $n=800$ & 0.026 & 0.023 & 0.021 & 0.025 & & 0.030 & 0.026 & 0.030 & 0.028 \\\addlinespace
      {ASE}& $n=200$ & 0.056 & 0.046 & 0.042 & 0.051 & & 0.061 & 0.052 & 0.057 & 0.057 \\
            & $n=400$ & 0.039 & 0.031 & 0.029 & 0.036 & & 0.043 & 0.036 & 0.040 & 0.039 \\
            & $n=800$ & 0.027 & 0.022 & 0.020 & 0.025 & &0.030 & 0.025 & 0.029 & 0.027 \\\addlinespace
      {CP}& $n=200$ & 0.934 & 0.944 & 0.911 & 0.954 & & 0.960 & 0.943 & 0.954 & 0.965 \\
            & $n=400$ & 0.882 & 0.943 & 0.864 & 0.919 & & 0.948 & 0.950 & 0.947 & 0.954 \\
            & $n=800$ & 0.729 & 0.936 & 0.724 & 0.802 & & 0.943 & 0.940 & 0.937 & 0.942 \\
        \hline
    \end{tabular}%
\end{table*}
\end{center}

\begin{center}
\begin{table*}[h]%
\caption{Simulation results on $\hat{R}^{-1}(p)$.\label{tab1}}
 \begin{tabular}{lcccccccccc}
       \hline\hline
       \\
    \multicolumn{2}{l}{Setting 1} & \multicolumn{4}{c}{GLM}       &       & \multicolumn{4}{c}{RCS} \\
    \multicolumn{2}{c}{} & $p=0.2$ & $p=0.3$ & $p=0.4$ & $p=0.5$ &       & $p=0.2$ & $p=0.3$ & $p=0.4$ & $p=0.5$ \\
    \hline
    \\
    \multicolumn{2}{l}{TRUE} & 0.173 & 0.387 & 0.589 & 0.744 &       & 0.173 & 0.387 & 0.589 & 0.744 \\\addlinespace
    Bias& $n=200$ & 0.006 & -0.027 & -0.035 & -0.024 &       & 0.027 & 0.007 & -0.019 & -0.031 \\
          & $n=400$ & 0.012 & -0.013 & -0.028 & -0.022 &       & 0.011 & 0.021 & -0.003 & -0.018 \\
          & $n=800$ & 0.010 & -0.013 & -0.026 & -0.020 &       & -0.001 & 0.016 & 0.000 & -0.013 \\\addlinespace
    ESE& $n=200$ & 0.089 & 0.097 & 0.091 & 0.091 &       & 0.101 & 0.111 & 0.101 & 0.087 \\
          & $n=400$ & 0.061 & 0.064 & 0.059 & 0.060 &       & 0.077 & 0.080 & 0.069 & 0.057 \\
          & $n=800$ & 0.046 & 0.047 & 0.045 & 0.045 &       & 0.058 & 0.059 & 0.052 & 0.042 \\\addlinespace
    ASE& $n=200$ & 0.077 & 0.095 & 0.094 & 0.086 &       & 0.089 & 0.108 & 0.107 & 0.094 \\
          & $n=400$ & 0.060 & 0.065 & 0.061 & 0.061 &       & 0.072 & 0.080 & 0.070 & 0.060 \\
          & $n=800$ & 0.044 & 0.046 & 0.043 & 0.043 &       & 0.055 & 0.057 & 0.049 & 0.042 \\\addlinespace
    CP& $n=200$ & 0.897 & 0.976 & 0.956 & 0.901 &       & 0.884 & 0.969 & 0.960 & 0.928 \\
          & $n=400$ & 0.934 & 0.969 & 0.925 & 0.913 &       & 0.903 & 0.948 & 0.954 & 0.946 \\
          & $n=800$ & 0.929 & 0.951 & 0.887 & 0.894 &       & 0.929 & 0.937 & 0.942 & 0.927 \\\addlinespace
    \multicolumn{11}{c}{} \\
    \multicolumn{2}{l}{Setting 2} & \multicolumn{4}{c}{GLM}       &       & \multicolumn{4}{c}{RCS} \\
    \multicolumn{2}{c}{} & $p=0.2$ & $p=0.3$ & $p=0.4$ & $p=0.5$ &       & $p=0.2$ & $p=0.3$ & $p=0.4$ & $p=0.5$ \\
    \hline 
    \\
    \multicolumn{2}{l}{TRUE} & 0.143 & 0.316 & 0.538 & 0.794 &       & 0.143 & 0.316 & 0.538 & 0.794 \\\addlinespace
    Bias& $n=200$ & 0.010 & 0.047 & 0.052 & -0.016 &       & 0.037 & 0.034 & 0.000 & -0.057 \\
          & $n=400$ & -0.024 & 0.028 & 0.062 & 0.011 &       & 0.007 & 0.011 & -0.001 & -0.018 \\
          & $n=800$ & -0.042 & 0.023 & 0.074 & 0.031 &       & -0.005 & 0.000 & -0.003 & 0.019 \\\addlinespace
    ESE& $n=200$ & 0.081 & 0.101 & 0.097 & 0.086 &       & 0.077 & 0.094 & 0.110 & 0.110 \\
          & $n=400$ & 0.061 & 0.079 & 0.074 & 0.069 &       & 0.058 & 0.070 & 0.092 & 0.093 \\
          & $n=800$ & 0.046 & 0.061 & 0.057 & 0.051 &       & 0.043 & 0.051 & 0.071 & 0.079 \\\addlinespace
    ASE& $n=200$ & 0.080 & 0.106 & 0.106 & 0.092 &       & 0.081 & 0.108 & 0.121 & 0.113 \\
          & $n=400$ & 0.059 & 0.083 & 0.080 & 0.071 &       & 0.058 & 0.080 & 0.098 & 0.094 \\
          & $n=800$ & 0.044 & 0.062 & 0.057 & 0.053 &       & 0.043 & 0.054 & 0.075 & 0.077 \\\addlinespace
    CP& $n=200$ & 0.923 & 0.931 & 0.988 & 0.940 &       & 0.920 & 0.959 & 0.988 & 0.932 \\
          & $n=400$ & 0.918 & 0.925 & 0.976 & 0.964 &       & 0.926 & 0.974 & 0.971 & 0.942 \\
          & $n=800$ & 0.906 & 0.904 & 0.814 & 0.980 &       & 0.952 & 0.971 & 0.948 & 0.931 \\
          \hline
    \end{tabular}%
\end{table*}
\end{center}

\section{Data Application}

Our example data consisted of 415 patients with cirrhosis who underwent transjugular intrahepatic portosystemic shunt (TIPS) procedures for refractory ascites or secondary prophylaxis of variceal bleeding \citep{vizzutti_mortality_2023}. This cohort was collected from the Istituto Mediterraneo per i Trapianti e Terapie ad Alta Specializzazione (ISMETT) in Palermo, Italy, between January 2007 and December 2019. Patient data were comprehensively recorded at the time of TIPS placement, covering a range of demographic and clinical variables. 

The outcome of interest was liver-related mortality, with orthotopic liver transplantation (OLT) and death from extrahepatic causes considered as competing events. Following the original paper \citep{vizzutti_mortality_2023}, we included age (years), alcoholic etiology, nonalcoholic steatohepatitis (NASH) etiology, international normalized ratio (INR), and creatinine levels as predictors of liver-related death. The median (first, third quartile) follow-up time was 10.6 (4.1, 30.1) months, with a maximum follow-up of 134 months. We set $\tau=60$ months
in the current analysis, where the overall cumulative incidence was 47\%.

\begin{figure*}
\centerline{\includegraphics[width=0.9\textwidth]{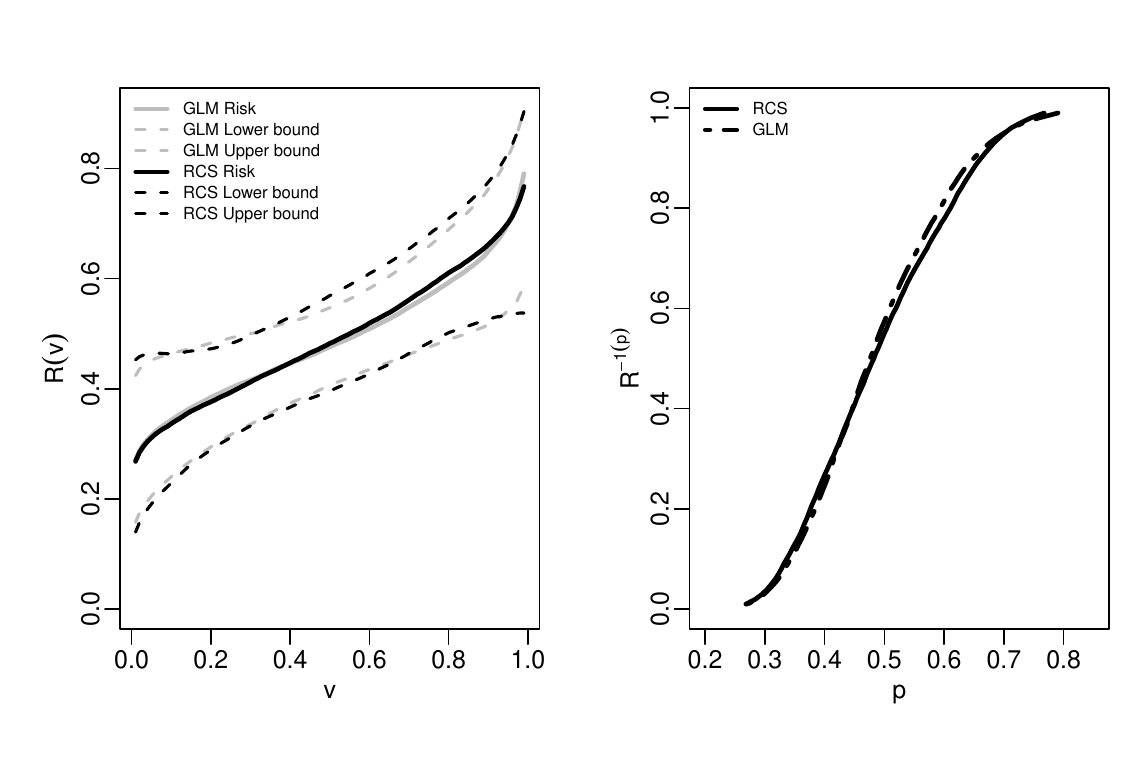}}
\caption{$\hat{R}(v)$ curve and $\hat{R}^{-1}(p)$ curve in data application.\label{fig3}}
\end{figure*}

Figure 3  represents the predictiveness curve estimate $\hat{R}(v)$ of liver‐related death, according to model-based risk score from Fine and Gray regression in cross-validations with 30 repetitions. The left panel for $\hat{R}(v)$ shows that the 5-year cumulative incidence of liver-related mortality ranged between 0.25-0.75 in this population when the quantiles of the model-based risk score increased from the lowest to the highest.
At the 20th percentile of the risk score, for example, the cumulative incidence of liver-related death was {0.38 (95\% CI 0.29 -- 0.47)}. At the 80th percentile of the risk score, the cumulative incidence of liver-related death increased to {0.61 (95\% CI 0.50 -- 0.71).} The GLM and RCS parameterization provided similar estimates for this dataset, with only slight variations.

In Figure 3, $\hat{R}^{-1}(p)$ curve shows the inverse relationship. The x-axis represents cumulative incidence probability thresholds, while the y-axis reflects the proportion of patients with 5-year cumulative incidence below each threshold in this population according to the model-based risk score. Both parameterizations showed a similar overall trend with slight deviations; therefore, we next interpret the RCS result. For example, we observed that around {25\%} participants of this population had cumulative incidence below 40\% ($\hat{R}^{-1}(0.4)$), while around 22\% subjects had cumulative incidence above 60\% ($1-\hat{R}^{-1}(0.6)$). Thus, around half of the participants fell into the intermediate risk brackets of $[40\%,60\%]$ in terms of their 5-year cumulative incidence of liver-related mortality.

\section{Discussion}
In this study, we develop estimation and inference procedures for predictiveness curve with competing risks prediction models. This approach allows for a detailed depiction of the risks level across the distribution of the predicted value, facilitating comprehensive understanding of the risk stratification performance in a population. Instead of assuming the regression model used for prediction as a true model, we allow it to be a working prediction model to enhance the applicability of the proposed method. We also use the repeated cross validation mechanism to mitigate the concern over estimation optimism, and flexible regression splines for good approximation of the relationship between the model-based risk score and the cumulative incidence function.

In this work, we have focused on a pre-specified prediction horizon $\tau$, which can be set at clinically relevant values in real applications. There are often commonly used horizons for specific disease fields, such as 5-year survival for cancer. One may also apply the proposed methods by setting $\tau$ at multiple values of interest. It would be desirable to further generalize the proposed method to provide a global evaluation with regard to different $\tau$s, such that the curve becomes a two-dimensional surface with $\tau$ and $v$. Additionally, while we focus on the overall predictiveness curve in the current work, it is of interest to evaluate conditional predictiveness curves given patient characteristics. These directions merit further research but are beyond the scope of the current paper.

\newpage
\bibliographystyle{plainnat}
\bibliography{mybib}

\begin{thebibliography}{20}
\providecommand{\natexlab}[1]{#1}
\providecommand{\url}[1]{\texttt{#1}}
\expandafter\ifx\csname urlstyle\endcsname\relax
  \providecommand{\doi}[1]{doi: #1}\else
  \providecommand{\doi}{doi: \begingroup \urlstyle{rm}\Url}\fi

\bibitem[Bellach et~al.(2019)Bellach, Kosorok, Rüschendorf, and
  Fine]{bellach_weighted_2019}
Anna Bellach, Michael~R. Kosorok, Ludger Rüschendorf, and Jason~P. Fine.
\newblock Weighted {NPMLE} for the {Subdistribution} of a {Competing} {Risk}.
\newblock \emph{Journal of the American Statistical Association}, 114\penalty0
  (525):\penalty0 259--270, 2019.
\newblock ISSN 0162-1459.
\newblock \doi{10.1080/01621459.2017.1401540}.

\bibitem[Bura and Gastwirth(2001)]{bura2001binary}
Efstathia Bura and Joseph~L Gastwirth.
\newblock The binary regression quantile plot: assessing the importance of
  predictors in binary regression visually.
\newblock \emph{Biometrical Journal: Journal of Mathematical Methods in
  Biosciences}, 43\penalty0 (1):\penalty0 5--21, 2001.

\bibitem[Ding et~al.(2021)Ding, Ning, and Li]{ding_evaluation_2021}
Maomao Ding, Jing Ning, and Ruosha Li.
\newblock Evaluation of competing risks prediction models using polytomous
  discrimination index.
\newblock \emph{Canadian Journal of Statistics}, 49\penalty0 (3):\penalty0
  731--753, September 2021.
\newblock ISSN 0319-5724, 1708-945X.
\newblock \doi{10.1002/cjs.11583}.
\newblock URL \url{https://onlinelibrary.wiley.com/doi/10.1002/cjs.11583}.

\bibitem[Escarela et~al.(2020)Escarela, Rodríguez, and
  Núñez‐Antonio]{escarela_copula_2020}
Gabriel Escarela, Carlos~Erwin Rodríguez, and Gabriel Núñez‐Antonio.
\newblock Copula modeling of receiver operating characteristic and
  predictiveness curves.
\newblock \emph{Statistics in Medicine}, 39\penalty0 (28):\penalty0 4252--4266,
  December 2020.
\newblock ISSN 0277-6715, 1097-0258.
\newblock \doi{10.1002/sim.8723}.
\newblock URL \url{https://onlinelibrary.wiley.com/doi/10.1002/sim.8723}.

\bibitem[Escarela et~al.(2023)Escarela, V{\'a}squez, Gonz{\'a}lez-Far{\'\i}as,
  and M{\'a}rquez-Urbina]{escarela2023copula}
Gabriel Escarela, Alejandro~Rom{\'a}n V{\'a}squez, Graciela
  Gonz{\'a}lez-Far{\'\i}as, and Jos{\'e}~Ulises M{\'a}rquez-Urbina.
\newblock Copula modeling for the estimation of measures of marker
  classification and predictiveness performance with survival outcomes.
\newblock \emph{Statistical Methods in Medical Research}, 32\penalty0
  (6):\penalty0 1203--1216, 2023.

\bibitem[Fine and Gray(1999)]{fine_proportional_1999}
Jason~P. Fine and Robert~J. Gray.
\newblock A {Proportional} {Hazards} {Model} for the {Subdistribution} of a
  {Competing} {Risk}.
\newblock \emph{Journal of the American Statistical Association}, 94\penalty0
  (446):\penalty0 496--509, June 1999.
\newblock ISSN 0162-1459, 1537-274X.
\newblock \doi{10.1080/01621459.1999.10474144}.
\newblock URL
  \url{http://www.tandfonline.com/doi/abs/10.1080/01621459.1999.10474144}.

\bibitem[Harrell(2001)]{harrell2001regression}
FE~Harrell.
\newblock Regression modeling strategies: with applications to linear models,
  logistic regression, and survival analysis, 2001.

\bibitem[{Harrell Jr}(2023)]{rmsmanual}
Frank~E {Harrell Jr}.
\newblock \emph{rms: Regression Modeling Strategies}, 2023.
\newblock URL \url{https://CRAN.R-project.org/package=rms}.
\newblock R package version 6.7-0.

\bibitem[Huang et~al.(2007)Huang, Sullivan~Pepe, and
  Feng]{huang_evaluating_2007}
Ying Huang, Margaret Sullivan~Pepe, and Ziding Feng.
\newblock Evaluating the {Predictiveness} of a {Continuous} {Marker}.
\newblock \emph{Biometrics}, 63\penalty0 (4):\penalty0 1181--1188, December
  2007.
\newblock ISSN 0006341X.
\newblock \doi{10.1111/j.1541-0420.2007.00814.x}.
\newblock URL
  \url{https://onlinelibrary.wiley.com/doi/10.1111/j.1541-0420.2007.00814.x}.

\bibitem[Jeong and Fine(2006)]{jeong2006direct}
Jong-Hyeon Jeong and Jason Fine.
\newblock Direct parametric inference for the cumulative incidence function.
\newblock \emph{Journal of the Royal Statistical Society Series C: Applied
  Statistics}, 55\penalty0 (2):\penalty0 187--200, 2006.

\bibitem[Jin et~al.(2001)Jin, Ying, and Wei]{jin2001simple}
Zhezhen Jin, Zhiliang Ying, and LJ~Wei.
\newblock A simple resampling method by perturbing the minimand.
\newblock \emph{Biometrika}, 88\penalty0 (2):\penalty0 381--390, 2001.

\bibitem[Klein and Andersen(2005)]{klein_regression_2005}
John~P. Klein and Per~Kragh Andersen.
\newblock Regression modeling of competing risks data based on pseudovalues of
  the cumulative incidence function.
\newblock \emph{Biometrics}, 61\penalty0 (1):\penalty0 223--229, March 2005.
\newblock ISSN 0006-341X.
\newblock \doi{10.1111/j.0006-341X.2005.031209.x}.

\bibitem[Li et~al.(2011)Li, Tian, and Wei]{li_estimating_2011}
Yi~Li, Lu~Tian, and Lee-Jen Wei.
\newblock Estimating {Subject}-{Specific} {Dependent} {Competing} {Risk}
  {Profile} with {Censored} {Event} {Time} {Observations}.
\newblock \emph{Biometrics}, 67\penalty0 (2):\penalty0 427--435, June 2011.
\newblock ISSN 0006341X.
\newblock \doi{10.1111/j.1541-0420.2010.01456.x}.
\newblock URL
  \url{https://onlinelibrary.wiley.com/doi/10.1111/j.1541-0420.2010.01456.x}.

\bibitem[Moskowitz and Pepe(2004)]{moskowitz_quantifying_2004}
C.~S. Moskowitz and M.~S. Pepe.
\newblock Quantifying and comparing the predictive accuracy of continuous
  prognostic factors for binary outcomes.
\newblock \emph{Biostatistics}, 5\penalty0 (1):\penalty0 113--127, January
  2004.
\newblock ISSN 1465-4644, 1468-4357.
\newblock \doi{10.1093/biostatistics/5.1.113}.
\newblock URL
  \url{https://academic.oup.com/biostatistics/article-lookup/doi/10.1093/biostatistics/5.1.113}.

\bibitem[Pepe et~al.(2007)Pepe, Feng, Huang, Longton, Prentice, Thompson, and
  Zheng]{pepe_integrating_2007}
M.~S. Pepe, Z.~Feng, Y.~Huang, G.~Longton, R.~Prentice, I.~M. Thompson, and
  Y.~Zheng.
\newblock Integrating the {Predictiveness} of a {Marker} with {Its}
  {Performance} as a {Classifier}.
\newblock \emph{American Journal of Epidemiology}, 167\penalty0 (3):\penalty0
  362--368, November 2007.
\newblock ISSN 0002-9262, 1476-6256.
\newblock \doi{10.1093/aje/kwm305}.
\newblock URL
  \url{https://academic.oup.com/aje/article-lookup/doi/10.1093/aje/kwm305}.

\bibitem[Sachs and Zhou(2013)]{sachs_partial_2013}
Michael~C. Sachs and Xiao‐Hua Zhou.
\newblock Partial summary measures of the predictiveness curve.
\newblock \emph{Biometrical Journal}, 55\penalty0 (4):\penalty0 589--602, July
  2013.
\newblock ISSN 0323-3847, 1521-4036.
\newblock \doi{10.1002/bimj.201200146}.
\newblock URL \url{https://onlinelibrary.wiley.com/doi/10.1002/bimj.201200146}.

\bibitem[Scheike et~al.(2008)Scheike, Zhang, and Gerds]{scheike2008predicting}
Thomas~H Scheike, Mei-Jie Zhang, and Thomas~A Gerds.
\newblock Predicting cumulative incidence probability by direct binomial
  regression.
\newblock \emph{Biometrika}, 95\penalty0 (1):\penalty0 205--220, 2008.

\bibitem[Stone(1986)]{stone1986generalized}
Charles~J Stone.
\newblock [generalized additive models]: comment.
\newblock \emph{Statistical Science}, 1\penalty0 (3):\penalty0 312--314, 1986.

\bibitem[Viallon and Latouche(2011)]{viallon2011discrimination}
Vivian Viallon and Aur{\'e}lien Latouche.
\newblock Discrimination measures for survival outcomes: connection between the
  auc and the predictiveness curve.
\newblock \emph{Biometrical Journal}, 53\penalty0 (2):\penalty0 217--236, 2011.

\bibitem[Vizzutti et~al.(2023)Vizzutti, Celsa, Calvaruso, Enea, Battaglia,
  Turco, Senzolo, Nardelli, Miraglia, Roccarina, Campani, Saltini, Caporali,
  Indulti, Gitto, Zanetto, Di~Maria, Bianchini, Pecchini, Aspite,
  Di~Bonaventura, Citone, Guasconi, Di~Benedetto, Arena, Fanelli, Maruzzelli,
  Riggio, Burra, Colecchia, Villa, Marra, Cammà, and
  Schepis]{vizzutti_mortality_2023}
Francesco Vizzutti, Ciro Celsa, Vincenza Calvaruso, Marco Enea, Salvatore
  Battaglia, Laura Turco, Marco Senzolo, Silvia Nardelli, Roberto Miraglia,
  Davide Roccarina, Claudia Campani, Dario Saltini, Cristian Caporali, Federica
  Indulti, Stefano Gitto, Alberto Zanetto, Gabriele Di~Maria, Marcello
  Bianchini, Maddalena Pecchini, Silvia Aspite, Chiara Di~Bonaventura, Michele
  Citone, Tomas Guasconi, Fabrizio Di~Benedetto, Umberto Arena, Fabrizio
  Fanelli, Luigi Maruzzelli, Oliviero Riggio, Patrizia Burra, Antonio
  Colecchia, Erica Villa, Fabio Marra, Calogero Cammà, and Filippo Schepis.
\newblock Mortality after transjugular intrahepatic portosystemic shunt in
  older adult patients with cirrhosis: {A} validated prediction model.
\newblock \emph{Hepatology}, 77\penalty0 (2):\penalty0 476--488, February 2023.
\newblock ISSN 0270-9139.
\newblock \doi{10.1002/hep.32704}.
\newblock URL \url{https://journals.lww.com/10.1002/hep.32704}.

\end{thebibliography}



\end{document}